# Data fusion techniques for fault diagnosis of industrial machines: a survey

Amir Eshaghi Chaleshtori[1], Abdollah aghaie[2*]

[1] ph.D. student of industrial engineering, K.N.Toosi University of Technology, Iran, Tehran; amir.isaqi@gmail.com

[2] Professor of industrial engineering, K.N.Toosi University of Technology, Iran, aaghaie@kntu.ac.ir

*Corresponding author: Abdollah Aghaie

**ABSTRACT**

In the Engineering discipline, predictive maintenance techniques play an essential role in improving system safety and reliability of industrial machines. Due to the adoption of crucial and emerging detection techniques and big data analytics tools, data fusion approaches are gaining popularity. This article thoroughly reviews the recent progress of data fusion techniques in predictive maintenance, focusing on their applications in machinery fault diagnosis. In this review, the primary objective is to classify existing literature and to report the latest research and directions to help researchers and professionals to acquire a clear understanding of the thematic area. This paper first summarizes fundamental data-fusion strategies for fault diagnosis. Then, a comprehensive investigation of the different levels of data fusion was conducted on fault diagnosis of industrial machines. In conclusion, a discussion of data fusion-based fault diagnosis challenges, opportunities, and future trends are presented.

## 1. INTRODUCTION

Industrial machines are becoming increasingly complex and specialized, increasing investment and machinery costs. In addition, the machines are in production lines and the production of Just-In-Time nowadays. However, if one component of a machine fails, the entire string, or even another company that depends on the produced workpiece, may become idle. Therefore, as much as possible, it is desirable to minimize machine downtime caused by unscheduled maintenance intervals and breakdowns. Moreover, it is essential to identify failures in advance to reduce failure costs. Given the adverse effects caused by the impending shortcomings, a great demand to find a method to solve this problem in the applications are observed, for instance, in the case of high-speed inverters [1], roller failures [2], the quad tank process [3], DC motors [4], bearings [5], the running gear in high-speed trains [6], and electrical drive systems [7].

Fortunately, the volume of data extracted from industrial machines has grown exponentially due to the plethora of detection and sensing technologies [8]. If data be processed and analyzed, valuable information and knowledge from the health status of machines are obtained, but based on massive data capture and customization of the manufacturing process, the ability to predict machine failures at a specific time in the future is a significant challenge in this field [9]. Therefore, the growing complexity of industrial machines and the amount of available information stimulate the derivation of fusion methods to meet this challenge [10]. It is a technology that combines data to achieve greater consistency and communication with original raw data that are mainly uncertain, imprecise, inconsistent, and contradictory. Merging or fusing is a fundamental process of combining or integrating different types of data that human typically performs this process using their brain automatically. Researchers, scientists, and industrialists have tried to mimic the ability of the human brain to merge data and information to create intelligence technologies. Researchers have proposed several techniques, approaches, and models to simulate the functioning of this process. Still, there is no unified definition to define data fusion, information fusion, approach fusion, and model fusion. Authors in [10] reviewed previous data and information fusion purposes and classified data and information fusion based on synonyms.

Nevertheless, data fusion or information fusion terms can be defined as the process of integrating [11,12] combining [13] a set of data or information from one or more sources. Various data fusion methods have been devised in different application areas. Data fusion is widely used in wireless sensor networks, image processing, radar systems, object tracking, target identification, intrusion detection, situation assessment, and condition monitoring [14].

This paper performs a literature review on data fusion methods for fault diagnosis of industrial machines, Identifies the streamlining of search in the literature, and delineates the research agenda for future research. The rest of the article's structure follows section 2 surveys the data fusion strategies. Section 3 presents the analysis of the recent related papers4, discusses the results, and identifies the existing research gaps, and Section 5 concludes this research.

## 2. SURVEY OF DATA FUSION STRATEGIES

There are two general types of Information fusion technologies: hard and soft [15, 16]. The distances and angles between sensors and their spatial arrangement are studied in the former. The latter refers to the development and improvement of algorithms employed in fusion strategies to assure accurate decision results, which also informs the present study. The

"soft" aspect of multi-sensory fusion in fault diagnosis has been studied at three strategic levels: fusion at the data level, fusion at the feature level, and fusion at the decision level [17].

In data-level fusion, data from each sensor or each different sensor type are aligned to make unified and integrated sensor data for further processing. This integrated sensor data will be considered the primary signal for further processing. This centralized fusion approach is theoretically the most accurate way to fuse data, assuming that the association and correlation can be performed correctly. In addition, this approach requires that raw data be transmitted from the sensors to a central processing facility or computer. Figure (1) shows the data-level fusion strategy used in machinery fault diagnosis.

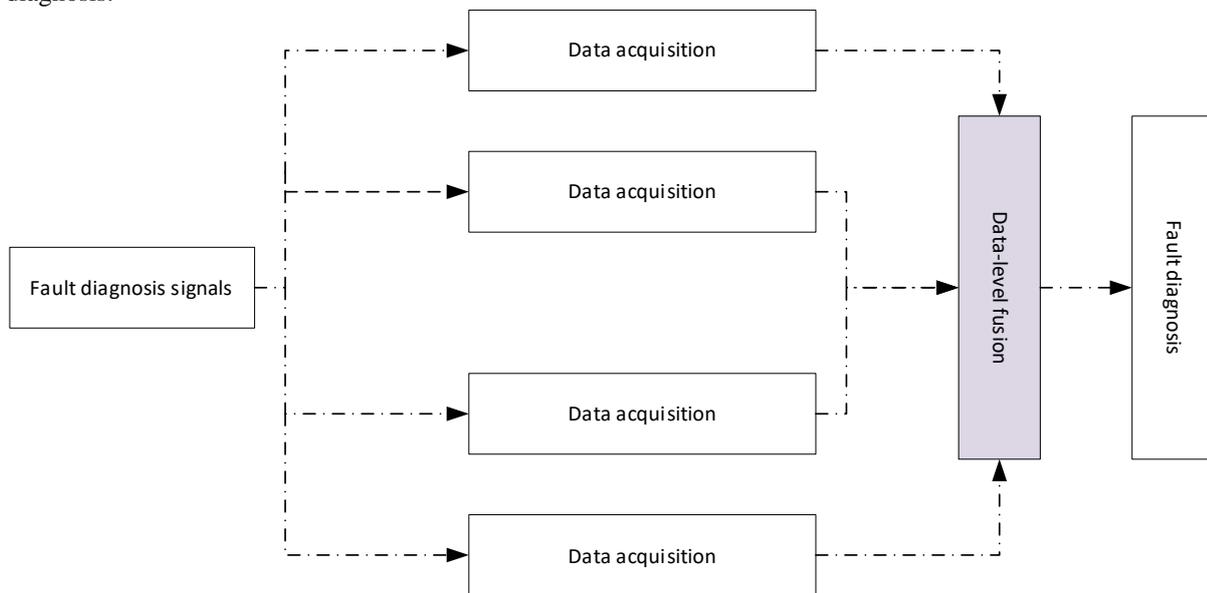

**Figure 1. Data-level fusion paradigm**

As shown in figure (2), the second architecture for fusion is feature-level fusion, in which each sensor performs single-source health estimation, producing a state of feature vector from each sensor. Depending on several sensors, these estimates are entered for a data integrating process to obtain a fused state vector. Note that the data alignment and association/correlation functions still need to be conducted. Feature fusion architectures reduce the unnecessary communications between the sensors and the integrating processors because the sensor data are compressed in a representative state vector. In addition, the association/correlation process is conceptually more straightforward than that performed to merge data layers; in general, feature-level fusion is not as accurate as data-level fusion because there is an information loss between the sensors and the fusion process. In particular, the original data contains information on the signal quality, which is only approximated by its associated covariance matrix.

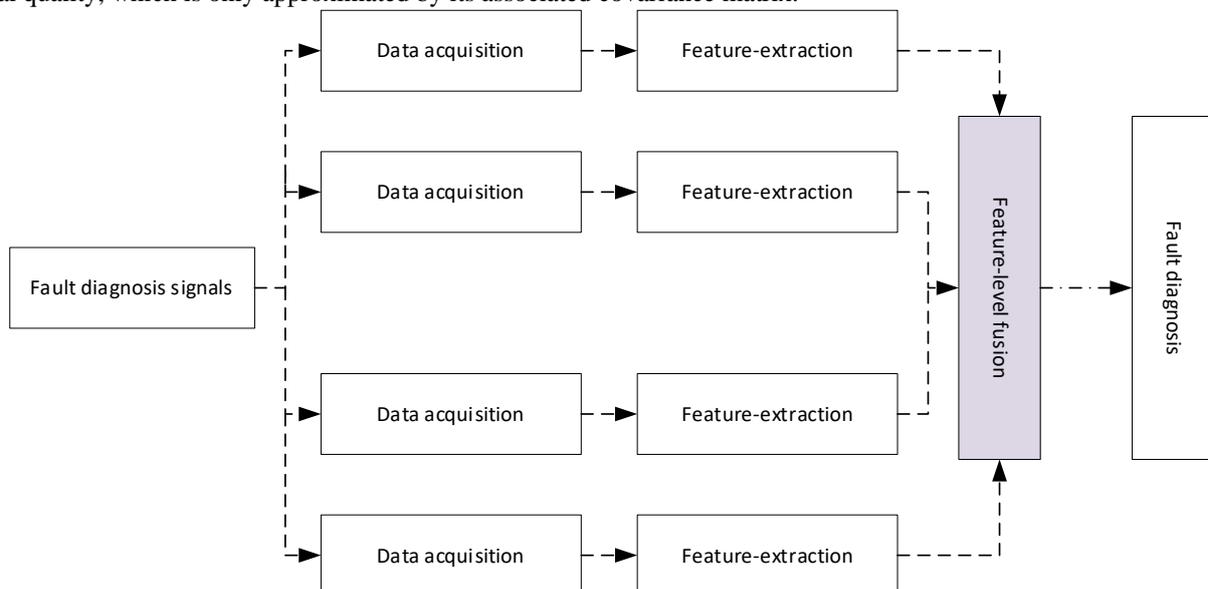

**Figure 2. Feature-level fusion paradigm**

Finally, the third architecture is decision-level fusion, illustrated in figure (3). Each sensor performs an identity declaration process in this architecture based on its data. Each sensor converts the observed target's attributes into a preliminary statement of the target's identity. Again, this can be done using a feature extraction/identity approach involving neural networks or other feature-based pattern recognition techniques. The identity declarations provided by the individual sensors are combined using decision-level fusion techniques such as classical inference, Bayesian inference, weighted decision methods, or Dempster–Shafer's theory of evidence.

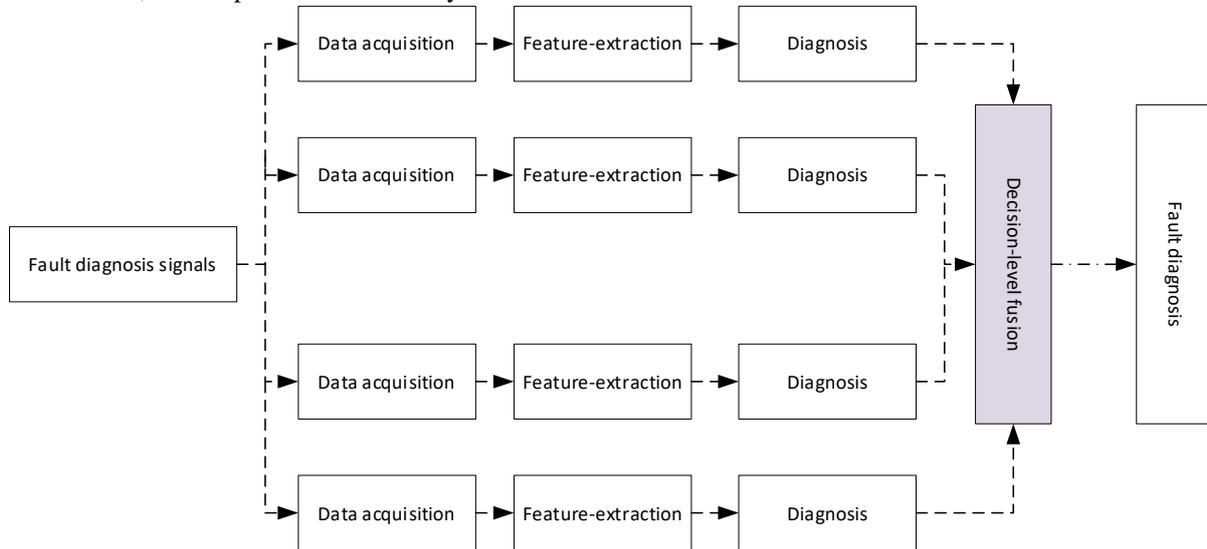

**Figure 3. Decision-level fusion paradigm**

The selection between these architectures for a particular application is also a system engineering problem that depends on issues such as available communication bandwidth, sensor characteristics, available computational resources, and other issues. There is no universal architecture applicable to all situations or applications.

### 3. ANALYSIS OF THE LITERATURE REVIEW

As mentioned before, the pipeline of typical data-driven based fault diagnosis techniques consists of three steps [18, 19]: first, signal preprocessing is performed to reduce noise and normalize multi-source fault signals; second, feature extraction\selection is performed from the preprocessed signals to reflect the health status of the machines. Third, pattern recognition is used by inputting the extracted defect features into a classifier to predict defect types.

In fusion at the data level, raw data is directly fused from different sensors. This strategy's strength is that it can utilize existing source data and overcome information loss [20]. In [20, 21], this fusion strategy has been applied for machinery fault detection.

Feature-level fusion is the process of integrating several original features into more meaningful features. In this fusion strategy, the original data may be contaminated by environmental noise, affecting the features' quality. Several signal decomposition methods can be employed to reduce the effect of noise, including wavelet analysis [22, 23], empirical mode decomposition (EMD) [24, 25], and frequency analysis [26]. Following decomposition and reconstruction [26, 27], the signal is decomposed to reduce noise influence and then decomposed again to extract fault features [28]. Despite the benefits of these methods, they have a few limitations: (I) They are not precise when applied to high-frequency decomposition; and (II) When removing noise, they distort the fault characteristics information. Stochastic resonance can enhance gains of periodic fault features with noise energy to determine the fault characteristic frequency. However, stochastic resonance cannot affect the total low-frequency interference [29, 30]. Moreover, the structural parameters of the resonance system can have a significant effect on fault diagnosis results. Several signal filtering methods can overcome these limitations, including high pass filtering [31], adaptive filtering [32], fractional filtering [33], and Kalman modifications [34], which analyze spectrum signals and diagnose faults to extract the actual frequency band. Another Challenge in feature-level fusion strategy is that many features extracted from signals can cover vast information. However, redundant information can also be immersed in it, reducing the accuracy of the model; increase complexity and computation time. Therefore, extracting useful information from the high-dimensional feature set is essential while eliminating redundant information. Fusion of higher dimensional features defined directly by data fusion techniques takes longer, leading to overfitting and the curse of dimensionality issues. By eliminating unnecessary information and noise, dimensionality reduction techniques can combine high-dimensional features into significant lower-dimensional features in machinery fault diagnosis. Selecting relevant feature subsets before merging is key to reducing complexity and overfitting. The literature proposes several feature selection methods to select essential features. Traditionally, filter-based [35], wrapper-based [36] and embedded methods [37] are used for feature selection. Thanks to the feature extraction methods, principal component analysis (PCA) is a commonly used traditional linear dimensionality reduction technique that accounts for the maximum variation in the data [38, 39]. A fault diagnosis and noise filtering in the data may be effectively performed with Independent Component Analysis (ICA) [40]. Moreover, Mahalanobis Distance (MD) has several ways to monitor health and identify various damage stages in naturally progressing defects [41]. However, in [42], a method for transformer fault classification using Factor Analysis (FA) was described. For fault detection and identification in a centrifugal compressor, [43] used Canonical Correlation Analysis (CCA) (also known as Canonical Variates Analysis). The idea behind linear discriminant analysis (LDA) is that the projection direction is determined by maximizing the distance between the data clusters commonly used to distinguish faults [44]. Even though the above methods are easy to implement and have shown effectiveness, they still have some drawbacks: 1) manual feature extraction from noisy signals requires much prior knowledge; and 2) the fault diagnosis process is isolated, likely to lose coupling information, causing low accuracy and weak generalization. Therefore, deep learning, as a branch of the machine learning method, provides a more comprehensive solution to the above problems. It has also received great attention and applications in fault diagnosis. It can learn features without requiring profound expert knowledge by analyzing the original input. Meanwhile, deep representative features can be extracted from a large amount of data with the multi-level structure of deep learning. Deep learning can extract non-linear features and has good generalizing abilities; it is an important branch of machine learning that can be applied to intelligent failure prediction. Several methods of deep learning are commonly used, including autoencoder (AE) based models, deep belief networks (DBN) models, convolutional neural networks (CNN) models, and recurrent neural networks (RNN) [45-48]. The challenges of intelligent fault diagnosis are that it is difficult to achieve higher performance and extract more features with single-mode samples and individual deep learning models. Therefore, combining different models and multi-modal data is necessary to obtain more robust and accurate fault diagnosis results. In addition to accuracy, cost calculation is also an essential part of applying industrial fault diagnostic models. As a rule of thumb, feature extraction is the most time-consuming step. A comparison between different classic CNN-based feature extractors is presented, such as Lent [49], Alexei [50], Googlebot [51], Magnet [52], and Reset [53]. A common theme of these approaches is that the more sophisticated the CNN model, the bigger the network parameters, which require more processing power and hardware. Therefore, various strategies are investigated to improve the soft structure and parameter tuning to save computation costs and economically increase iteration speed. On the other hand, the decision-level fusion method combines different models, followed by a weighted fusion decision for the outputs of those models [54]. Many decision-level-based diagnostic techniques have been proposed to diagnose failures accurately and can be classified in statistical and machine learning methods. Hidden Markov Model (HMM) and the Bayesian method are well-known statistical methods for analyzing failures [55, 56]. However, the validity of statistical models is limited when hypotheses such as normality, linearity, or independence of variables are not satisfied. Since these limitations exist, machine learning techniques have gained more attention recently for fault diagnosis, including artificial neural networks (ANNs), support vector machines (SVMs), decision trees, random forests, and k-nearest neighbors (KNNs), and extreme learning machines (ELMs). [19, 57-61].

Moreover, studies have shown that combining multiple learners could achieve more accurate fault diagnosis results [56, 62, 63]. So, ensemble learning methods have been applied, which increase the stability of obtained results by combining base learning models with fusion strategies after training a group of base learners. In the generation phase, subsets can be subdivided based on instances or features applied in fault diagnosis [64-66]. Furthermore, recent studies have shown that implementing these methods generally requires the artisanal extraction of the characteristics of the fault signal in the time domain, the frequency domain, the time-frequency domain, and the nonlinear domain.

Due to these challenges, the application of deep learning (DL) algorithms for machinery fault diagnosis research has grown exponentially due to its high accuracy, efficient data representation, automatic feature extraction, and selection [19]. By stacking multi-level structures, DL algorithms provide complex objective functions and improve model generalization by discovering the relationships between variables [18, 19, 67]. In decision-level machinery fault diagnosis, DL algorithms are commonly used, such as auto-encoders (AE) [68-70], convolutional neural networks (CNN) [71-73], deep belief networks (DBN) [17, 74], deep residual networks (DRN) [75, 76], and recursive neural networks (RNN) or its variant long short time memory networks (LSTM) [77, 78]. CNN and its improved versions have attracted the most attention in diagnosing machine failures of all the DL algorithms. As a network-improvement technique, CNN is best used in conjunction with other ML/DL methods, such as ELM [73], DBM [74], RNN [78], and DRN [75, 76]. Although CNN can satisfactorily diagnose machine failures, its network training is often challenging because a large training sample is needed to avoid overfitting test samples [79]. Therefore, it is necessary to use data augmentation techniques if the amount of raw data is not large enough.

## 4. RESULTS AND DISCUSSION

With the ever-faster development of sensor and communication technologies, technology ecosystems are recommended to organize items from multiple sources to obtain sensor data [80, 81]. Hence, more than one sensor is installed in different locations and directions to maximize the capture of underlying fault information from industrial machines. The various strategies for fusing this multi-sensor data have advantages and disadvantages, as shown in Table 1. Therefore, selecting an optimal fusion strategy for a practical application remains challenging.

Although the loss of information in data-level fusion is minimal, this type of fusion has limited exploration, mainly because the raw data provided by each sensor is typically large, and the amount of noise is significant. Therefore, theoretical explanations and guidelines for dealing with different data sources, transmission paths, or sampling strategies for a good fusion are rarely discussed.

With moderate information loss, feature-level fusion can compress large datasets into small datasets with representative information [82]. However, different features need to be extracted and chosen for merging in various troubleshooting tasks. Manual feature extraction is still based on subjective experience knowledge of domination and human labor [83]. Despite the highest information loss, decision-level fusion is practically easy to implement [82]. In addition, deep learning has a powerful ability to learn functions directly from raw data and can largely overcome information leakage. Multi-sensory fusion integrated with deep understanding has received increasing attention in the areas of fault diagnosis. However, we need a reasonable voting-merge strategy to get collaborative fault diagnosis results with high precision and good stability in practical tasks. Average Voting and Weighted Voting are two popular merging strategies. Nevertheless, neither is the idea that the average vote assumes that each basic diagnostic model is equally important and makes an equal contribution. However, some sensor signals can change dramatically when an error occurs, while others change very little.

Table 1. pros and cons of data fusion strategies

| Fusion Strategy | Level | Advantage | Disadvantage |
| --- | --- | --- | --- |
| Data-level | Low | Retain all of the information | Lack of interpretability for records registration. |
| Feature-level | Middle | Compress high-dimensional information units into small-scale information | Need area know-how and hand exertions for characteristic extraction. |
| Decision-level | High | Good practically and interpretability | Sensitive to voting fusion rules. |

Moreover, IoT-based multi-source sensing data fusion technology can be divided into three phases: data acquisition, fusion modeling, and application of fusion algorithms. The data collection phase significantly impacts the industrial application of mechanical fault diagnosis and prediction technology. The problems and challenges are summarized below.

(1) Sensor selection issues: When presenting real mechanical devices, it may be necessary to select different sets of sensors for data collection in different environments. Also, for the same type of sensors, the industry standards for sensors from other manufacturers may differ, resulting in various degrees of error. Also, the sensor group may vary when applied to different mechanical equipment. Can be developed, and these sensors can be meaningfully combined for data collection according to actual conditions.

(2) Frequency of data collection: If the frequency of data collection is too high, data redundancy will occur. On the other hand, if the sampling frequency is too low, the data obtained cannot be used reliably for fault diagnosis. The system is usually operated continuously and the data set generated is relatively large. In a further development, one of the main goals is to establish a data acquisition frequency standard for different sensors and to integrate the original data for the first time.

Following are some of the main challenges faced in multi-source data fusion based on the status of the development of fusion models and fusion algorithms:

(1) The fusion model is not uniform: In the field of mechanical fault diagnosis and prediction, there is no unified framework for dealing with the various scenarios of mechanical fault diagnosis. Existing fusion models are usually specific to a particular type of device. Future research should focus on designing and building a unified fusion framework for diagnosing faults in mechanical equipment.

(2) There is much noise in the original data resulting from uncontrollable environmental factors during the data collection. The decision result is inaccurate if the original data is used directly for feature extraction and data fusion. When presented with raw data, combining the fusion model with the fusion algorithm is necessary to select an appropriate data preprocessing method. In the future development process, it would be beneficial to establish a set of data preprocessing techniques for different sensors used in fault diagnosis and prediction, specifically for mechanical equipment.

(3) Long running time: When fusion algorithms based on deep learning are used for training, the running time required to find suitable hyperparameters is typically long, and sometimes overfitting occurs. As a result, research on fusion algorithms mainly focuses on feature and decision-level fusion. Moreover, there are very few algorithms that deal with data-level fusion. Thus, further development of data-level fusion algorithms is required in the future development process.

## 5. CONCLUSION

This document has provided a comprehensive review of the literature on data fusion for fault diagnosis of industrial machines. First, we provide basic knowledge about data fusion. Then the works in this document are reviewed to assess the advantages and disadvantages of fusion strategies. On the other hand, we have carefully reviewed the current literature based on the decomposed fusion level in industrial machine fault diagnosis. Based on our survey, we have further specified the number of open issues and suggested future research directions that require further investigation. This study provides a concise and comprehensive reference for researchers and practitioners in the field of data fusion for industrial machine fault diagnosis.